\documentclass{emulateapj}
\usepackage{graphicx}

\def\lea{\mathrel{<\kern-1.0em\lower0.9ex\hbox{$\sim$}}}
\def\gea{\mathrel{>\kern-1.0em\lower0.9ex\hbox{$\sim$}}}

\slugcomment{To appear in AJ }
\shorttitle{Star Clusters in M51} 
\shortauthors{Lee, Chandar \& Whitmore}

\begin{document}

\title{Properties of Resolved Star Clusters in M51}

\author{\sc Myung Gyoon Lee\altaffilmark{1,2}, Rupali Chandar\altaffilmark{3}, and Bradley C.\ Whitmore\altaffilmark{3}}
\altaffiltext{1}{Astronomy Program, SEES, Seoul National University, Seoul 151-742, Korea}
\altaffiltext{2}{Visiting Investigator at the Department of Terrestrial Magnetism, Carnegie Institution
of Washington, 5241 Broad Branch Road, N.W., Washington, DC 20015}
\altaffiltext{3}{Space Telescope Science Institute, 3700 San Martin Dr., Baltimore, MD 21218}

\email{mglee@astrog.snu.ac.kr, rupali@stsci.edu, whitmore@stsci.edu}

\begin{abstract}
We present a study of compact star clusters in the nearby pair of
interacting galaxies NGC 5194/95 (M51), based on multifilter {\it
Hubble Space Telescope\/} WFPC2 archival images.  We have detected
$\sim$400 isolated, resolved clusters in five \textit{HST} WFPC2 fields of
the two galaxy system.  Due to our requirement that the clusters be
detected based only on their morphology, which results in the
selection of relatively isolated objects, we estimate that we are
missing the majority (by a factor 4--6) of clusters younger than
$\sim$10~Myr due to the extreme crowding in the spiral arms and
star-forming regions.  Hence we focus on the cluster population older
than 10~Myr.  An age distribution of the detected clusters shows a
broad peak between 100--500~Myr, which is consistent with the
crossing times of the companion galaxy NGC~5195 through the NGC~5194
disk estimated in both single and multiple-passage dynamical models.
We estimate that the peak contains $\sim$2.2--2.5 times more clusters
than expected from a constant rate of cluster formation over this time
interval.  While there is also evidence for individual peaks near
100~Myr and 500~Myr in the cluster age distribution (consistent with
the predictions of multiple-passage models), this result requires
verification.  We estimate the effective radii of our sample clusters
and find a median value of $\sim$3--4~pc.  Additionally, we see
correlations of (increasing) cluster size with cluster mass (with a
best fit slope of $0.14\pm0.03$) at the $\sim4\sigma$ level, and with
cluster age ($0.06\pm0.02$) at the $3\sigma$ level.  Finally, we
report for the first time the discovery of faint, extended star
clusters in the companion, NGC~5195, an SB0 galaxy.  These have red
[$(V-I)>1.0$] colors, effective radii $>7$~pc, and are scattered over
the disk of NGC~5195.  Our results indicate that NGC~5195 is therefore
the third known barred lenticular galaxy to have formed so-called
``faint fuzzy'' star clusters.
\end{abstract}


\section{INTRODUCTION}

Star cluster systems track the star formation histories of their host
galaxies.  While these star clusters have been relatively well studied
in nearby starburst and strongly interacting galaxies (e.g., Meurer et~al.\ 1995; 
Whitmore et~al.\ 1999; also see compilations in Whitmore
2003; Larsen 2004), there remains much to learn concerning the
properties of cluster systems in more ``normal'' environments, such as
those found in spiral galaxies.

The M51 system (NGC~5194/95) is a famous, nearby interacting pair of
galaxies.  NGC~5194 is a grand design, Milky-Way-like (Sbc) spiral,
and its close companion NGC~5195 is a dwarf barred spiral of
early-type (SB0).  NGC~5194 is almost face-on (with the eastern side
tilted by an estimated 20~degrees; \citet{tul74b}), and shows several
remarkable features in addition to the two grand-design spiral arms.
These include faint extended tidal features deviating from the spiral
arms \citep{bur78}, and a very long (90~kpc) HI tidal tail ($5\times
10^8~M_\odot$) extending to the west \citep{rot90}.
While the optical spectrum of NGC~5194 is typical for Sbc galaxies,
that of NGC~5195 shows strong Balmer absorption lines indicating star
formation activity in the recent past \citep{ken98}.

Given the crowded conditions in spiral disks due to on-going star
formation, the excellent resolution provided by the {\it Hubble Space
Telescope\/} (\textit{HST}) is required to study the cluster populations in
external spirals.  Prior to the beginning of this study, the \textit{HST}
archive contained five distinct WFPC2 field pointings of the M51
system.  Bik et~al.\ (2003) and Bastian et~al.\ (2005) have recently
made detailed studies using observations from two of these pointings,
including all available filters.  Some NICMOS imaging is available for
the very center of NGC~5194.

In this paper, we present a study of star clusters in the M51 system,
based on the analysis of \textit{HST} WFPC2 archive images of the five fields
covering roughly 60\% of the optically bright regions of the two
galaxies.  We do not repeat the Bastian et~al.\ (2005) work here, but
rather use different detection techniques in order to minimize
contamination from blends in our final cluster sample, and include
three additional WFPC2 fields to make a more general study of the M51
system.  Because of this selection based on morphology and isolation,
there is a strong bias in our sample against the young clusters (e.g.,
$\lea10$--20~Myr), which tend to form in very crowded regions.

This paper is organized as follows: Section~2 describes the data,
reduction, cluster selection and photometry.  \S3 analyses the
properties of the detected clusters including age, reddening, mass,
size, and spatial distribution, and a discussion of the main results
is presented in \S4.  Finally, a summary is provided in \S5.
\newpage

\section{DATA, CLUSTER SELECTION, AND PHOTOMETRY}

\subsection{Data and Reduction}

\begin{figure}
%\epsscale{1.}
\centering
\includegraphics[width=4in]{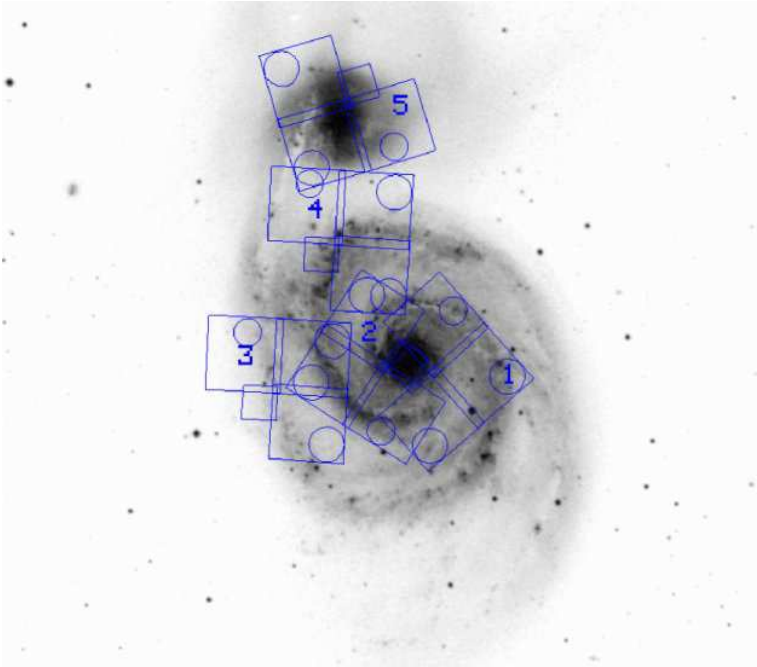}
\caption{The five $HST$ WFPC2 field pointings used in this study are
overlaid on a $15\arcmin \times 15 \arcmin$ Digitized Sky Survey image of 
the M51 system.
\label{figure:fov}}
\end{figure}

We have examined five archival \textit{HST} WFPC2 pointings in M51.  All have
at least two filter observations.  As with any study based on
archival data, we are limited by the availability of filters, exposure
times, and field pointings.  
Figure~\ref{figure:fov} shows the locations of the five fields
overlaid on a $\sim15^{\prime}\times15^{\prime}$ Digitized Sky Survey
image.  The fields are numbered as in Table~1, which contains general
data properties, such as the field identification, central right
ascension (RA) and declination (DEC), and the number and exposures in
the various filters.  Surveyed fields in this study cover different
environments, from the nuclear region to both arm and inter-arm
regions, as well as some portions between NGC~5194 and its barred
lenticular companion NGC~5195.  The center of NGC~5194 is included in
Fields~1 and~2 (on the PC chip) and the center of NGC~5195 is almost
centered in Field~5.

Field~1 has \textit{UBVI} and $H\alpha$ imaging, while Fields~2, 3, and~4
have \textit{BVI} observations (Field~5 is only imaged in \textit{VI}).  Field~2
additionally has an $R$ band image, and partial overlap with $U$ band
observations.  All data was requested from the \textit{HST} Archive with
``on-the-fly'' calibrations, which automatically use the best
reference files for calibration.  The WFPC2 pipeline steps include:
bad pixel masking, A/D correction, bias and dark subtraction, and flat
field correction.

Multiple individual images in a given filter were combined using the
COMBINE task in STSDAS with the CRREJECT option in order to eliminate
cosmic rays (CRs) and increase the signal-to-noise ratio.  The IRAF
task COSMICRAYS was then run on the combined images to eliminate
remaining CRs and individual hot pixels.  In order to correct for the
well known geometric distortion in the WFPC2 CCDs, as well as the
34th row problem, each image was multiplied by a correction image.

We used the F656N filter observations to determine whether a source in
Field~1 has associated $H\alpha$ emission (the presence of $H\alpha$
helps break the degeneracy between age and reddening which exists in
analysis based only on broad-band colors).  The $H\alpha$ flux was
determined by using a scaled version of the $I$ band image of Field~1
to subtract continuum flux and reveal the sites of ionized gas
emission.  We estimated the continuum level from the I band images
from several bright, isolated objects.  Although the offset in
wavelength between $H\alpha$ and the I band is not optimal, this
technique is sufficient for our purposes, since $H\alpha$ emission
drops off rapidly over the first several Myr in the lifetime of a
cluster.

We adopt a distance to M51 of $8.4\pm0.6$~Mpc (distance modulus,
$(m-M)_0=29.62$), determined from the planetary nebula luminosity
function \citep{fel97}. %(Feldmeier et al. 1997).
The corresponding linear scale is 40.7 pc~arcsec$^{-1}$.  The \textit{HST}
WFPC2 has pixel sizes of $0.1\arcsec$ in the three Wide Field CCDs,
and $0.0455\arcsec$ in the Planetary Camera CCD.  At the assumed
distance, the four M51 fields (1--4) cover a range of (projected)
galactocentric distances from the nucleus to $350\arcsec$ (14.2~kpc),
and a total area of $\sim19.3$ square arcminutes ($115~\mbox{kpc}^2$).
The foreground reddening toward M51 is low \citep[$E_{B-V}=0.035$;][]{sch98}.

\subsection{Cluster Selection}

\begin{figure}
\centering
%\epsscale{0.8}
\includegraphics*[width=5in]{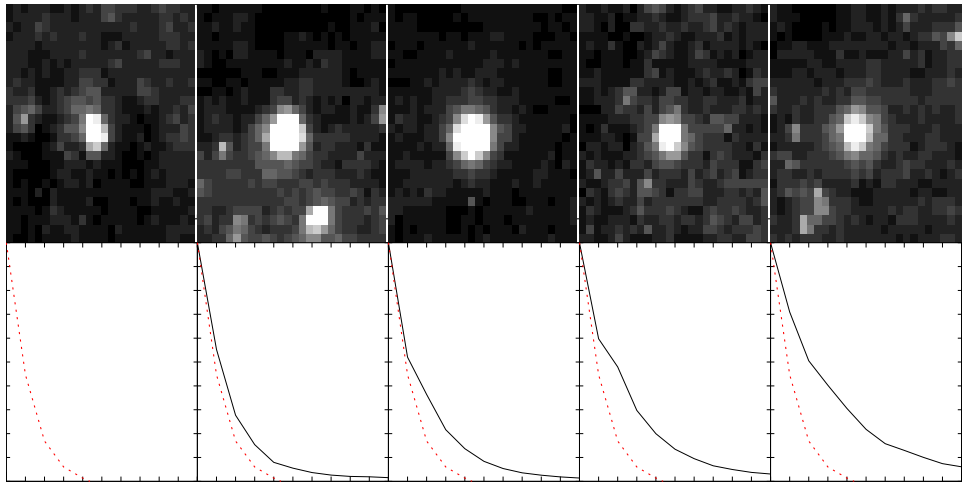}
\caption{Figure showing the radial profiles of representative star
clusters in M51 compared with a stellar profile.  The top left panel
shows a star, and the other (top) panels show example $V$ band images of
four clusters from our sample.  Each image is $2.5\arcsec \times
2.5\arcsec$.  Below each object image we have plotted (in arcseconds)
its radial profile (solid line), with that of the star (dotted line)
shown for comparison.
\label{figure:profiles}}
\end{figure}

Images of nearby spiral galaxies contain individual stars (both
foreground and residing in the host galaxy itself), background
galaxies, star clusters, and numerous observational blends, where
several objects are superposed.  While a visual inspection of these
blends clearly indicates the presence of two or more overlapping
objects, they are often detected as a single object by the finding
algorithm.  Hence, detecting star clusters in the complex, crowded
environments of nearby spirals is not trivial.  However, within at
least 10~Mpc, compact clusters can generally be detected based on
their morphological properties in \textit{HST} images.
One goal of this work is to correlate cluster sizes with other
properties; therefore we take a conservative approach and study only
relatively isolated and well resolved clusters in M51.  To accomplish
this, we make use of the information in radial profiles of detected
objects, plus other morphological criteria, in order to separate
clusters from stars, background galaxies, and blends.
Figure~\ref{figure:profiles} shows $V$ band images and radial profiles
of one star and four clusters in M51, as an example.  We note that
compact clusters with $r_{eff}\lea1.2$~pc ($\sim0.6$~pc) at the
distance of M51 are not clearly resolved on the WF (PC) CCDs.

In order to identify star clusters using morphological information, we
used SEXTRACTOR \citep{ber96} to detect objects in the $V$ band images.
A threshold of $4\sigma$ was used to avoid large numbers of detections
of very faint objects (generally noise artifacts) in our non-uniform
background, crowded fields.  A number of morphological parameters were
output for each detected object.  In addition, point spread function
(PSF) fitting was performed on each object.  Isolated stars were
chosen automatically using size, shape, and neighbor information, and
used to create a PSF.  This PSF was then fit to all cataloged objects
using the IRAF task ALLSTAR,
which outputs a goodness of fit statistic, $\chi^2$ and a measure of
object sharpness.  Cluster candidates were then chosen based on their
FWHM ($\geq1.8$~pix), ellipticity ($\leq 0.2$), $\chi^2$ ($\geq 1.2$),
and sharpness ($\geq0.18$) parameters.  The FWHM distribution for the
entire catalog of detected objects shows a peak near 1.5~pixels, which
is the FWHM of the (undersampled) PSF of the WFPC2.  Although the FWHM
distribution dips sharply around 1.8~pixels, it is a continuous
distribution.  Our experiments showed that if the FWHM threshold was
decreased below 1.8~pixels, a number of apparently unresolved objects
were added to our catalog of cluster candidates, and if we increased
the FWHM criterion to 2.0~pixels, we started to lose a small fraction
of clearly resolved objects from our cluster catalog.  In general,
however, the largest variations in the selected cluster candidates
with FWHM came from the very crowded, star-forming regions, which we
avoided (as described below).

The FWHM and sharpness
parameters allowed us to separate extended sources from stars, and the
ellipticity criterion eliminates a handful of background galaxies, and
more importantly, a large number of blends consisting of superposed
stars (measured FWHMs of these multiple sources are typically larger
than those of individual stars, and thus overlap the measurements for
star clusters).

Visual inspection of the initial cluster candidates showed a large
number of objects residing in crowded OB associations and spiral arm
regions, which were clearly superpositions of several objects.
Because these have similar parameters to isolated clusters, we
experimented with a number of techniques for automatically eliminating
these spurious cluster candidates.  After extensive testing, we
eliminated objects which had two or more cluster candidates within 10
(22) pixels for the WF (PC) CCDs.  This eliminated most of the obvious
blends in crowded regions.  Remaining blends were then rejected by a
final visual inspection.  Although it is not clear what fraction of
these detected blends are actual clusters, we are almost certainly
excluding a large number of real clusters from our catalog.  In
\S4.2 we discuss how our technique of excluding objects detected as
cluster candidates in very crowded regions impacts the age
distribution of the overall cluster population.

This prescription resulted in the detection of 392 isolated clusters
in M51, with $V$ magnitudes down to 23.5.  While this is clearly not a
complete list of clusters in this two-galaxy system, it is a
relatively clean cluster sample.  Thus, by restricting our sample to
isolated, resolved clusters, our sample is mostly insensitive to small
changes in the (FWHM, ellipticity, sharpness and $\chi^2$) selection
criteria.

\begin{figure}
\centering
%\epsscale{0.75}
\includegraphics[width=2.5in]{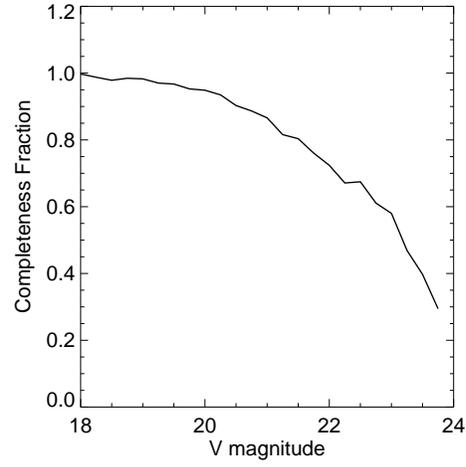}
\caption{An average $V$ band completeness curve as determined from artificial
cluster experiments (described in text) is shown.
\label{figure:complete}}
\end{figure}

Artificial cluster experiments were performed by adding artificial
clusters (generated from the ADDSTAR task in IRAF, where instead of
stars, clusters were selected) to our images, which were then rerun
through SEXTRACTOR and the automated portion of our detection
algorithm.  These 'fake' clusters were added in groups of 50 in
randomly placed positions on each chip, and then detected and
re-photometered.  Thus we make the assumption that as long as a
cluster makes it through the automated pipeline, it would also be
retained during the final visual inspection phase (which was used
primarily to weed out blends).  In Figure~\ref{figure:complete} we
show a $V$ band completeness function for Field~1.  Averaged over all
environments available in Field~1, Figure~\ref{figure:complete} shows
that the 50\% completeness limit is reached at $V\sim23$.  However,
the typical S/N for clusters as faint as $V=23$ is too low to make an
accurate estimate of the cluster size via fitting an analytic function
(as described in \S3.4).  Therefore, in the following analysis, we
generally adopt a detection limit of $V\sim22$, where clusters are
detected with a S/N of 30 or higher.

\subsection{Comparison of Cluster Candidates with Previous Works}

\begin{figure}
\centering
%\epsscale{1.}
\includegraphics[width=3.5in]{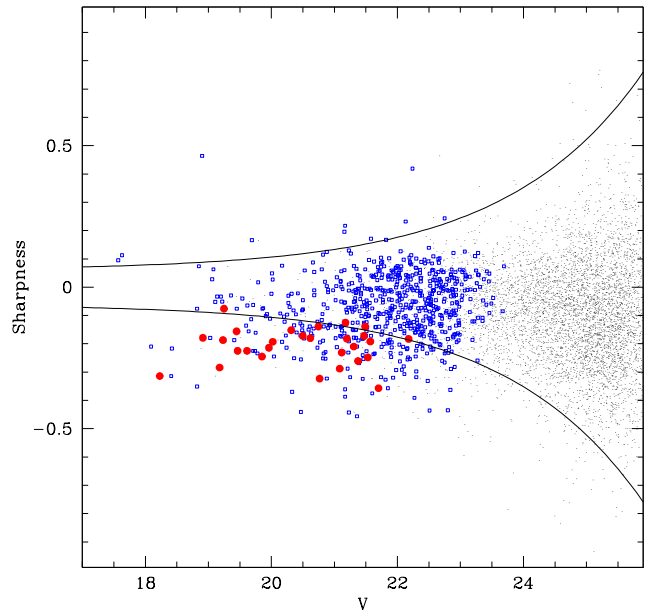}
\caption{Plot showing the sharpness vs.\ $V$ magnitude results from
PSF fitting for our clusters (circles), Bik et al.\ (2003) 
cluster candidates (squares), and
all detected objects (dots) in the WF Chip~2 in Field~2 of M51.  The
two solid lines represent $2.5\sigma$ envelopes.
Objects inside the envelope are considered to be
unresolved point sources, while those below the lower envelope are 
considered to be extended sources.  Note that all of our clusters are 
below or along the lower envelope, while most of the Bik et al.\ (2003) 
cluster candidates are between the envelopes.
\label{figure:hstphot}}
\end{figure}

Bik et~al.\ (2003) presented a list of 877 cluster candidates in the
WF2 CCD of Field~2 (461 of these are brighter than $V=22$), while
we only have 30 clusters in the same region---a significantly smaller
number.  While this is partly by design since we focus on relatively
isolated clusters which are brighter than $V\sim22$, this is still a
large difference.  To understand this large discrepancy between the
two studies, we made a sanity check on our cluster selection
techniques using a PSF fitting algorithm.
In Figure~\ref{figure:hstphot} we plot sharpness vs.\ $V$~magnitude for
all the sources detected in the WF CCDs of Field~2 (dots), including
our clusters (filled circles) and Bik et~al.\ (2003) cluster candidates
(open squares) from the same image.  The solid lines represent
$2.5~\sigma$ boundaries which were defined from the data with positive
sharpness (where there is expected to be no extended objects) using
the exponential functions given in Dolphin \& Kennicutt (2002).  This
figure establishes that all the candidates in our sample are below (or
along) the lower boundary, where clusters are expected to reside,
while most of the cluster candidates in the Bik et~al.\ (2003) sample
are above the lower boundary (in the regime occupied by point
sources).  Hence, the Bik et~al.\ sample is potentially contaminated by
a large number of stars.  However, this issue was revisited in Bastian
et~al.\ (2005), where the authors also use size measurements to
distinguish point sources (presumably stars) from resolved star
clusters, and find that restricting their sample to resolved objects
does not change the main conclusions of their study.  We note that due to our
selection criteria, the most compact star clusters (which are barely
resolved) are likely missing from our sample.

Many of the objects which fall in the regime of resolved star clusters
in Figure~\ref{figure:hstphot}, but were not selected by us as such,
are either fainter than we probe in our study (size estimates become
very difficult at these faint magnitudes), or are in very crowded
regions.  We can use the additional cluster candidates selected by the
PSF fitting algorithm (and seen in Figure~\ref{figure:hstphot}) to
estimate how many clusters we are missing from our sample.  By
counting the total number of cluster candidates (below the $2.5\sigma$
envelope) in this Figure which are brighter than $V=22$, and comparing
with the total number of clusters we have selected (filled circles),
we estimate that our current ``clean'' cluster sample underestimates
the number of resolved clusters by a factor $\sim$4--6 in this
magnitude range.  In \S3.3 we use integrated colors to assess the
age distribution of these rejected cluster candidates.

\subsection{Cluster Photometry}

\subsubsection{Photometry}

Aperture photometry was obtained using the PHOT task in DAOPHOT
\citep{ste94}.  We used a small aperture radius ($r=3$ pixels) to
determine cluster colors, in order to minimize contamination due to
neighboring sources and to reduce the impact of uncertainties in the
background determination.  However, there is a significant fraction of
light outside this radius, which varies based on how extended an
object is, and which directly impacts the total $V$~magnitude
measurements (and hence cluster mass estimates).
	
In order to measure the aperture corrections for our cluster
candidates, we identified relatively isolated clusters on the PC CCD
and WF CCDs.  The final aperture corrections of $-0.39$~mag (PC) and
$-0.31$~mag (WF CCDs) were added to the cluster $V$~magnitudes.  While
aperture corrections are dependent on intrinsic object size, making
absolute magnitudes somewhat uncertain (see \S2.4.2), we note that
the {\it colors\/} of the clusters are largely unaffected.

The following steps were used to transform measured WFPC2 instrumental
magnitudes to Johnson-Cousins $U$, $B$, $V$, $R$ and $I$~magnitudes:
{\it (i)}~the instrumental magnitudes were corrected for the
charge-transfer efficiency (CTE) loss, using the prescription given by
\citet{dol00a}\footnote{see
http://www.noao.edu/staff/dolphin/wfpc2\_calib/ for updated calibrated
information.}; {\it (ii)}~the corrected instrumental magnitudes were
converted to standard Johnson-Cousins $U$, $B$, $V$, and $I$~magnitudes.  
Using Equation~8 and Table~7 of \citet{hol95}, the
magnitudes were derived iteratively using WFPC2 observations in two
filters, with all zeropoints substituted from \citet{dol00a}.

\subsubsection{Comparison with Previous Work}

Here, we compare our photometric measurements of the objects in common
with previous studies \citep{lar00a, lam02, bik03}, including
both ground-based and $HST$ photometry.
Among the 69 cluster candidates discovered in ground-based images by
\citet{lar00a}, we find 13 objects in common.
Our $(B-V)$ colors agree well, with $\Delta (B-V)= -0.02\pm0.08$.
However, our total $V$~magnitudes are typically fainter by $\Delta V=
+0.42\pm0.32$ \citep[where $\Delta$ means this study minus][]{lar00a}.

\citet{lar00a} also used the \textit{HST} observations of Field~2 to
compare with his ground-based photometry for some clusters. 
He was able to locate ten cluster candidates in the \textit{HST}
image.  Of these ten, we find that four are clearly blends.
A comparison between this work and that of \citet{lar00a} shows that
the \textit{HST} photometry are in good agreement---the mean difference in
the $V$~band magnitudes is 0.002 mag, and the mean difference in color
$\Delta(B-V)$ is 0.048 mag, in the sense that our $(B-V)$ color is
slightly {\it redder\/} than that given in \citet{lar00a}.  This may
result from local background determinations and slight differences in
CTE corrections.  Overall however, we find that our photometry is in
very good agreement with the \textit{HST} photometry from \citet{lar00a}, and
conclude that the large uncertainties reported in the previous
paragraph are due to the ground-based measurements rather than a
problem with our \textit{HST} photometry 
\citep[as also seen from Table~2 in][]{lar00a}.

\citet{bik03} presented photometry of the objects in the image of the
WF2 CCD of Field~2.  The mean differences in photometry for 28 objects
in common with our study are: $\Delta V=-0.194\pm0.083$, $\Delta
(B-V)=0.059\pm0.054$, and $\Delta (V-I)=-0.002\pm0.035$ where $\Delta$
means this study minus \citet{bik03}.  Again, there are some
systematic differences in the total magnitudes between the two works,
but the colors are in reasonably good agreement.

\section{Analysis}

\subsection{Cluster Age, Extinction, and Mass Derivation}

In order to determine the age and reddening ($E_{B-V}$) intrinsic to
each cluster, we compared the observed magnitudes with spectral energy
distributions (SED) derived from the theoretical evolutionary
synthesis models of Bruzual \& Charlot (2003; hereafter BC03).  The
models assume that the stars have a \citet{sal55} %Salpeter (1955)
initial-mass function (IMF) slope $\frac{dlogN}{dlogM}=-2.35$, with
lower mass cutoff $0.1~M_{\odot}$ and upper mass cutoff $125~M_{\odot}$.  
For each metallicity, the models span ages from 1~Myr to
15~Gyr.

These spectral synthesis models are available for a number of
metallicities; however, due to the well known
age-metallicity-reddening degeneracy in integrated cluster colors, we
need to adopt a specific value for the metallicity, since our dataset
is not sufficient to solve for all three parameters independently.
Observations of H~II regions in M51 establish that the current
metallicity of the gas is approximately solar 
\citep[e.g.,][]{dia91,hil97}, hence we adopt this value for the subsequent
analysis.  Tests show that the assumed metallicity has a negligible
effect on the derived ages and extinction values for younger stellar
populations ($\lea 1$~Gyr), but preferentially affects the ages
estimated for older clusters, where metallicity influences become more
pronounced than those of age in the integrated colors.

For clusters in Fields~1 and~2, where there are a minimum of four
broad-band observations, we fit the observed SEDs of the clusters with
the models using a standard $\chi^2$ minimization technique.  For each
BC03 model age, we compare the observed SED to the model, which was
reddened by $E_{B-V}$ values between 0.0 and 1.0 in steps of 0.02.
This range of reddening values appears reasonable, as previous studies
have found typical color excess values of 0.2 in the bulge of NGC~5194
\citep{lam02}.  The fit with a minimum value of $\chi^2$ was adopted
as the best fit age/$E_{B-V}$ combination.
We use the $H\alpha$ measurements as a ``knife-edge,'' in the sense
that if $H\alpha$ emission is measured (from the continuum subtracted
image) with S/N~=~5 or higher, then the $H\alpha$ is given a high weight
in the SED fitting, which effectively prefers a younger age.

For clusters in Fields~3 and~4, where there is no $U$~band imaging
available, it is not possible to unambiguously separate reddened young
clusters from ancient objects.  Therefore, we selected ancient cluster
candidates by requiring $V-I\geq0.8$ and $B-V\geq0.55$.  This results
in the selection of 37 clusters with colors similar to those of
Galactic globular clusters in NGC~5194.  However, a visual check shows
that three of those with only \textit{BVI} filter observations fall directly
in active star forming regions (i.e., in dust lanes).  These are
assumed to be reddened young objects rather than ancient star
clusters, giving a total of 34 ancient star cluster candidates in
NGC~5194.  The properties of these objects are studied in detail in
Chandar, Whitmore, \& Lee (2004), and the main conclusion is that we
find a lack of red (metal-rich) globular clusters in NGC~5194.

For the clusters where there are only \textit{BVI} observations it is
difficult to get a handle on the reddening for each cluster
individually, since changes in extinction mimic changes in age.
Instead, we adopt the technique described in Lamers et~al.\ (2002).  We
adopted the probability distribution of $E(B-V)$ found for clusters in
Field~1.  This results in a $\chi^2$ value at each model age for every
value of $E(B-V)$.  The reduced $\chi^2$ which results from the fit is
used to distinguish between accepted and rejected fits.  For the final
age estimate, we average the ages, weighted by the probability that
each value of $E(B-V)$ occurs. While the age and reddening values for 
any given cluster using this technique may be incorrect, {\it statistically\/} 
the cluster populations age and reddening distributions should be 
reasonably well represented.

Cluster masses were derived by combining the $M/L_V$ ratio of the best
fit (age) BC03 model with the measured $V$~band cluster luminosity,
foreground extinction, derived cluster extinction, and assumed
distance to M51.  Masses of the clusters range from $\log M/M_\odot
=2.8$ to 6.3, whereas the average Galactic globular cluster mass is
$\sim2 \times 10^5~M_\odot$ \citep{har91}.  The most massive cluster
in the sample is estimated to have $M=1.9 \times 10^6~M_\odot$.
Apparent mean masses of younger clusters are lower than those of older
clusters, although it should be kept in mind that this is largely due
to selection effects (e.g., the magnitude threshold removing the older,
low mass clusters).  The mass function of young clusters in M51
appears to include objects which are comparable in mass to typical
Milky Way globular clusters.

\begin{figure}
%\epsscale{1.}
\centering
\includegraphics[width=3in]{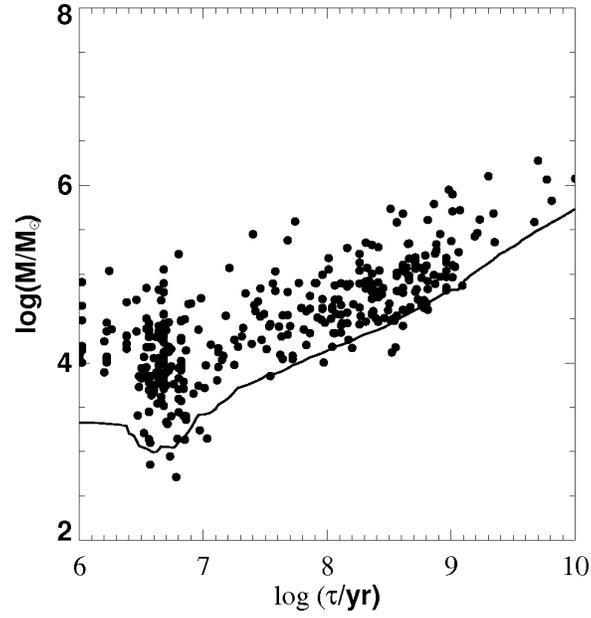}
\caption{Derived age vs.\ mass for our NGC~5194 cluster sample.  The solid
line shows how our detection limit of $V=22$ affects our ability to
detect star clusters as a function of age.  
\label{figure:agevsmass}}
\end{figure}

In Figure~\ref{figure:agevsmass} we show the derived
age versus mass plot for our M51 cluster sample.  The solid line shows
the fading with age for a cluster with $V=22$, based on the predictions
of the BC03 models.  Hence, our study cannot detect ancient clusters
($\gea$few~Gyr) with masses lower than $\sim10^5~M_{\odot}$.
The upper envelope in this figure can be represented by a linear line
with a slope of 0.4.  This trend is almost certainly due to a size-of-sample effect
(i.e.,  the total original number of clusters with ages in the range $9
\leq \mbox{log(age[yr])} \leq10$ is likely to be several orders of
magnitude larger than in the range $6 \leq \mbox{log(age[yr])} \leq
7$, resulting in the formation of clusters with larger masses in older
age bins).

There are several interesting features in
Figure~\ref{figure:agevsmass}.  First, we note the apparent ``gap''
between 10--30~Myr in the cluster distribution.  This is an artifact
common to age-dating techniques which rely on the comparison of
integrated colors with the predictions of evolutionary synthesis
models.  The predicted colors in this age range change rapidly, and
thus small photometric uncertainties in measured cluster colors make
it difficult to obtain best fit ages in this range.  A second
prominent feature in the age vs.\ mass diagram is an apparent
over-density of clusters with ages $\mbox{log(age[yr])}\lea7$ (even
despite the fact that we are missing a large number of such young
objects in our sample).  This over-density of clusters at very young
ages is likely not entirely due to a very recent, enhanced star
formation rate, but rather related to the rapid destruction of
clusters, as discussed in Fall (2004); Whitmore (2004); Fall, Chandar,
\& Whitmore (2005).  Finally, we note that there appears to be a
second over-density in the cluster population in the range $8 \lea
\mbox{log(age[yr])} \lea 9$.  We believe this feature {\it does\/}
represent an actual enhancement in the star formation {\it rate\/} in
the M51 system.  In \S4.3 we discuss and quantify this effect.

\subsection{Comparison of Ages Derived from UBVI filters vs.\ BVI Filters}

In the previous section, we described our techniques for deriving ages
and reddening values depending upon whether there are three or four
filters available.  Because only Field~1 has complete $U$~band
observations, here we discuss the limitations of the age derivation
from three filters, and the impact this may have on our conclusions
concerning the cluster age distribution.

\begin{figure}
\centering
%\epsscale{1.}
\includegraphics[width=3in]{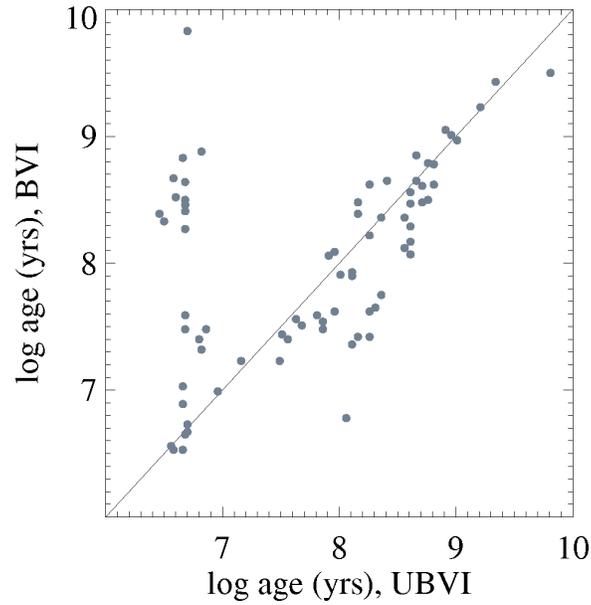}
\caption{Comparison of Field~1 ages estimated using \textit{UBVI} and \textit{BVI} SED fitting 
techniques. 
Only clusters with good fits are included.
\label{figure:agecomp}}
\end{figure}

Figure~\ref{figure:agecomp} shows clusters from Field~1 which have
reasonably well fit ages from both \textit{UBVI} and \textit{BVI} SED fitting.  The
solid line shows the one-to-one correspondence of ages.  Overall, we
find that there is good correlation between the ages estimated with
the two techniques.  The obvious exception to this are the 12~points
which have young age estimates based on \textit{UBVI} filters
($\mbox{log(age[yr])}<7.0$), but have significantly larger estimates
($\mbox{log(age[yr])}>8.0$) from \textit{BVI}.  As it turns out, these
objects are reddened young star clusters; their \textit{BVI} colors are
typical of older objects, but with the addition of the $U$~band flux,
the degeneracy between age and reddening can be broken.  We
statistically corrected the age distribution of our cluster population
for this bias in age estimates when only \textit{BVI} filters were available.

\subsection{Assessing Sample Bias Due to Selection}

In \S2.2 we described our selection of star cluster candidates based
on morphological criteria.  We found that a number of objects are
selected during automated detection in very crowded inner and spiral
arm regions.  A number of these are clearly superpositions of multiple
sources or blends.  While at least some of these objects are likely
star clusters, based on the resolution of the WFPC2 data it is not
possible for us to directly assess what fraction falls in this
category.  Therefore in this work, we focus on the properties of the
relatively isolated cluster population in M51.  However, in order to
draw any global conclusions concerning the M51 cluster population, we
must first quantify properties of clusters which may have been
excluded.

To accomplish this we studied Field~1, which has the most complete
filter selection available (\textit{UBVI}$H\alpha$).  We used our original
object catalogs to locate cluster candidates which were eliminated due
to crowding, and performed aperture photometry as was described in
\S3.1.  By comparing measured colors with evolutionary synthesis
model predictions, as we have done for the sample clusters, we find
that $\sim$80--90\% of the objects eliminated due to crowding have ages
younger than 10~Myr.  A visual inspection of the location of these
sources in the continuum subtracted $H\alpha$ image confirms that most
of these fall in regions where the emission from ionized gas is
strong, and hence in regions dominated by very young stellar
populations.  Therefore, we conclude that our current sample is
missing a large fraction (by factors 4--6 as quantified in \S2.3)
of very young clusters, but is not missing a significant population of
older objects.

\subsection{Cluster Sizes}

Intrinsic sizes for our entire cluster sample were measured using the
ISHAPE routine.  A detailed description of the code is given in
\citet{lar99}, along with the results of extensive performance
testing.  Essentially, ISHAPE measures intrinsic object sizes by
adopting an analytic model of the source and convolves this model with
a (user-supplied) point spread function (PSF), and then adjusts the
shape parameters until the the best match is obtained.  King model
profiles with concentration parameters of $c=30$ were convolved with a
PSF, and fit individually to each object.  The input PSF to this
algorithm is crucial.  Hence we created a PSF by hand-selecting stars
in the image, and then compared the results with those from a
subsampled TinyTim PSF (when the TinyTim PSF was used, convolution
with the WFPC2 diffusion kernel was implemented as recommended in the
ISHAPE manual).  We found that the size estimates from ISHAPE using
these two PSFs differed by less than 20\%.  Final measurements were
made using the TinyTim PSF. One PSF was
generated for the PC CCD, and one for the WF CCDs.  The size
measurements were made on the $V$~band images.

\begin{figure}
\centering
%\epsscale{1.}
\includegraphics[width=3.5in]{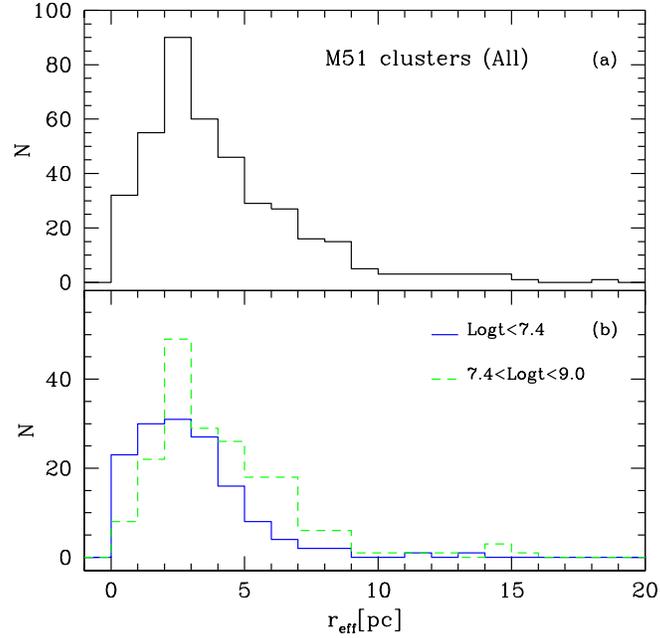}
\caption{Distribution of effective radii for our entire NGC~5194 cluster 
sample (top panel), and for two age groups (bottom panel).
\label{figure:sizehist}}
\end{figure}

Figure~\ref{figure:sizehist} displays the distribution of cluster
effective radii.  Our fixed size cut at a SEXTACTOR FWHM of 1.8~pixels
maps roughly to an ISHAPE FWHM measurement of 0.2~pixels; 0.2~pixels
is recommended in Larsen 1999 as a cutoff for separating resolved and
unresolved objects.  At the distance of M51, this FWHM corresponds to
0.8~pc or $r_{eff}$ of 1.2~pc.  In the current dataset our choice in
SEXTRACTOR FWHM cutoff is reflected by the rapid drop in measured
cluster sizes below $\sim$2~pc.  The mean size of our entire cluster
sample is $r_{eff}=3.7~\mbox{pc}$, and the median size is 3.1 pc.
This is very similar to the typical 3--4~pc found for Galactic
globular cluster effective radii \citep{har96}.  However, there
appears to be a ``tail'' in the size distribution, out to
significantly larger radii.  Given that clusters with size
measurements $\geq7$~pc have a large range of S/N values (from
30--150) which overlap those for the bulk of the more compact
clusters, we rule out that objects with large size measurements are
simply due to larger measurement errors. We have checked the location 
of the extended clusters in the WFPC2 images, and these also do not 
fall preferentially near the edges of a CCD.

\section{DISCUSSION}

\subsection{Correlation of Cluster Sizes with Age and Mass}

\begin{figure}
\centering
%\epsscale{1.}
\includegraphics[width=4in]{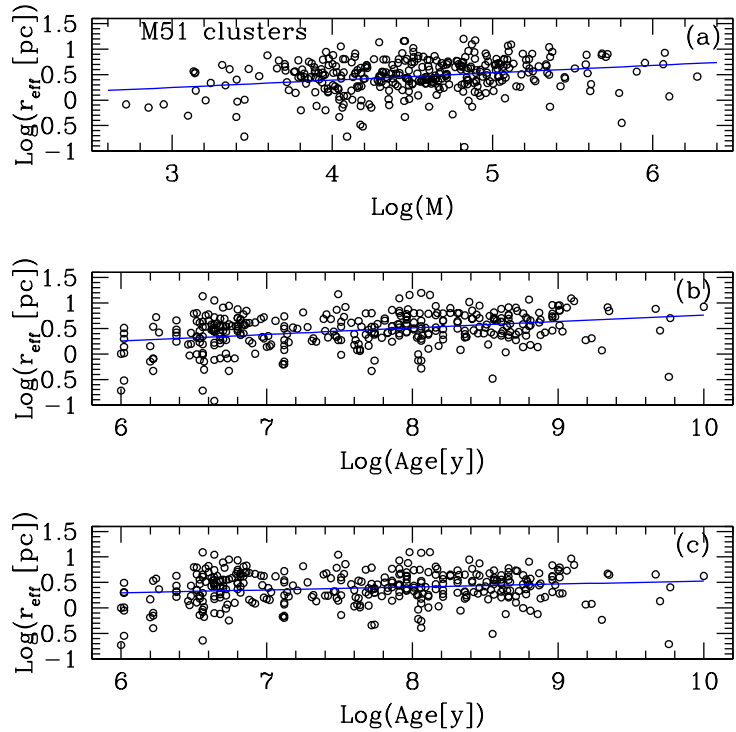}
\caption{The top panel plots the effective radius of our NGC~5194
cluster sample versus their derived (log) mass.  The middle panel
shows the effective radius versus estimated cluster age (log yrs) for
our NGC~5194 cluster sample.  In the bottom panel, we correct the
cluster size distribution for the derived size-mass relation, and
refit.  The solid lines represent the best linear fits,
and the slopes are $0.14\pm0.03$, $0.13\pm0.02$ and $0.06\pm0.02$ for
the three panels respectively.
\label{figure:sizes}}
\end{figure}

In general, it might be expected that if a cluster forms once its
parent molecular cloud reaches a critical density, the resulting
cluster would also reflect this density.  This would result in an
observed increase in the cluster size with cluster mass (with cluster
size increasing proportional to $M^{1/3}$).  In
Figure~\ref{figure:sizes} we plot the measured sizes for our NGC~5194
cluster sample as a function of derived mass (top panel) and age
(middle panel).  The solid lines in each panel represent the best
linear fit.  The top panel in Figure~\ref{figure:sizes} has a best fit
slope of $0.14\pm 0.03$.  This fit formally indicates a correlation at
the $\sim4\sigma$ level between cluster $r_{eff}$ and mass, albeit
with large scatter.  This is similar (within $\sim1.5\sigma$) to the
result found by Larsen (2004) for a large sample of young star
clusters in 18 nearby spirals using \textit{HST} WFPC2 imaging, and by Hunter
et~al.\ (2003) for clusters in the Large and Small Magellanic Clouds.
Bastian et~al.\ (2005) however, did not see this trend in their sample
of M51 clusters.  Despite the trend of increasing cluster size with
mass, the observed relation for clusters in M51 is significantly
shallower than predicted by a constant density relation.

In the middle panel of Figure~\ref{figure:sizes} we see a higher
fraction of very compact ($r_{eff}\sim1$ pc) {\it young\/} clusters in
our sample, relative to similarly compact clusters at older ages.  The
best linear fit to the data give a slope of $0.13\pm0.02$, implying a
trend of increasing cluster size with age.  However, since we found a
relationship between cluster size and mass, the observed size--age
relationship may simply reflect the fact that the observed mass limit
changes with age (such that more massive and larger clusters are
preferentially detected at older ages).  In order to establish whether
there is an intrinsic trend between cluster sizes and ages, we
subtracted out the derived cluster mass-size relation from the cluster
sizes, and refit the data (as shown in the bottom panel of
Figure~\ref{figure:sizes}).  This results in a linear fit of
$0.06\pm0.02$; a weaker trend than found before correcting for the
trend of increasing size with mass, but significant at the $3\sigma$
level.
\newpage

\subsection{Cluster Age and Spatial Distributions}

\begin{figure}
\centering
%\epsscale{1.}
\includegraphics[width=4in]{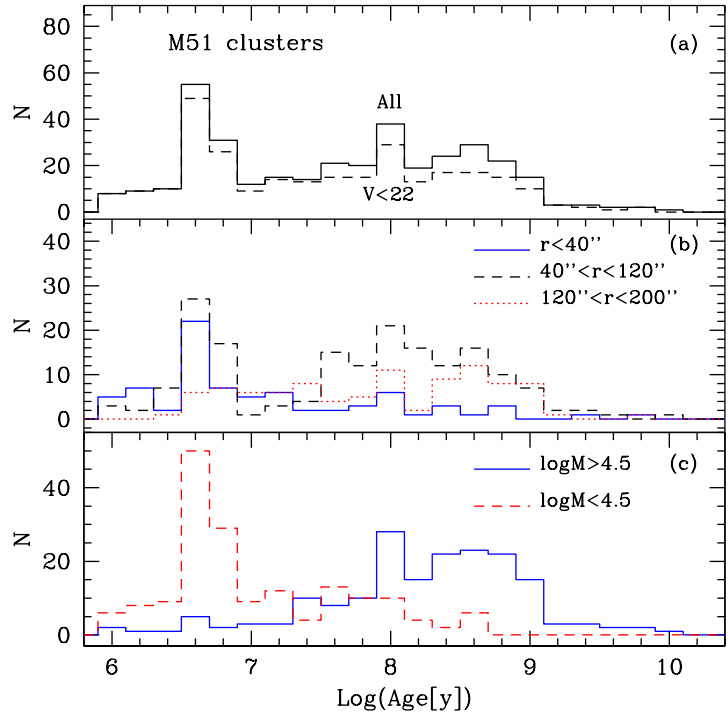}
\caption{
Age distributions of all sample clusters
in NGC~5194, as well as those brighter than $V=22$ (dashed line).  The middle 
panel shows the age distribution at different distances from the center
of NGC~5194, and the bottom
panel shows the age distribution for two different mass ranges.
\label{figure:dndt}}
\end{figure}

The age distribution for our NGC~5194 cluster sample is presented in
Figure~\ref{figure:dndt}.  The top panel in this figure shows several
features.  While NGC~5194 has been continuously forming star clusters,
there are at least two apparent peaks above the background at
$\mbox{log(age[yr])}6.6$ and a broad peak between 8.0 and 9.0 in log yrs.
These peaks are seen both in the entire cluster sample and in the
sample of bright clusters with $V\leq22$ mag where cluster detection
is reasonably complete.  In the bottom panel of
Figure~\ref{figure:dndt} we show the age distribution for clusters
with masses estimated above $3\times10^4~M_{\odot}$.  This sample is
expected to be reasonably complete out to $\sim$1~Gyr, and again shows
a broad peak in the number of clusters formed in M51 between
log(age[yr]) 8.0 and 9.0.  The number of clusters observed in the young peak
is significantly reduced in the bottom panel, since these are
primarily of lower mass; however recall that we are likely missing a
factor of 4--6 young clusters due to our selection criteria.

\begin{figure}
\centering
%\epsscale{1.}
\includegraphics[width=3.5in]{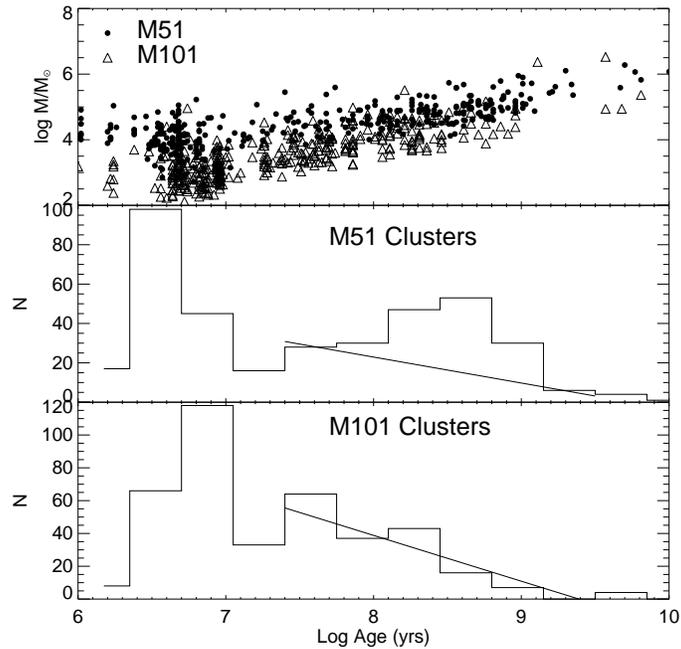}
\caption{The top panel in this figure shows the estimated age versus mass
for star clusters discussed in this paper for M51, and for a sample of
star clusters detected in a deep \textit{HST} WFPC2 image of M101 (Chandar
et~al., in preparation).  The second and third panels show the age 
distributions for M51 and M101 cluster systems, as well as a linear fit
between ages of $7.5 < \mbox{log(age[yr])} < 9.5$.  The plots show the
excess of intermediate age
clusters in M51 relative to those in M101 (which is believed to have had
a more quiescent formation history).
\label{figure:toymodel}}
\end{figure}

In order to quantify the enhancement in the cluster formation rate in
Figure~\ref{figure:dndt} (which is also clearly seen in
Figure~\ref{figure:agevsmass}), we
first synthesized a population of star clusters assuming a constant
cluster formation model (a description of these models are presented
in Whitmore 2004; Whitmore \& Chandar 2005), assigned a random
extinction, $E_{B-V}$ between 0.0 and 0.4, and assumed a photometric
uncertainty to each cluster which depends on the luminosity of the
cluster in a given bandpass.  The characteristic uncertainties were
determined empirically from the actual M51 cluster observations.  We
added a random correction to the magnitude of each cluster in each
filter, and then ran this synthesized cluster population through our
age-dating code, to account for systematic biases in the age-fitting.
These models show a linear decline in the cluster distribution with
age, with no apparent bumps.  However, since there can be additional
sources of uncertainty when comparing observations with models, as a
sanity check we also compared the M51 cluster age distribution to that
found for clusters detected in a single WFPC2 pointing of M101 (these
will be presented in Chandar et~al., in preparation), which has \textit{UBVI}
filter observations.  M101 is a relatively isolated galaxy with no
known companion, and hence we might naively expect it to have a
more quiescent cluster formation history then M51.
In the top panel of Figure~\ref{figure:toymodel} we compare the
results of the cluster populations in these two galaxies.
A histogram of the number of clusters as a function of age is shown
for M51 (middle panel) and M101 (bottom panel).  Note that we have
used broader age bins here than those used in Figure~\ref{figure:dndt}
(in order to be more robust against small number statistics), which
washes out the apparent bimodality between $7.6 < \mbox{log(age[yr])}
< 9.2$ (with apparent peaks at log (age) 8.0 and 8.6).  In both the
constant cluster formation model (not shown) and the M101 cluster
population, the number of clusters observed between $7\leq
\mbox{log(age[yr])} \leq 10$ decreases linearly, regardless of the bin
width that is used.  For M51 however, we clearly see an excess in the
number of clusters with ages between log(age[yr])~8 and~9.  The fact that
we see this contrast between derived properties for the {\it observed\/}
cluster populations in M101 and M51, using the same age-dating
technique, gives us additional confidence that the observed broad peak
in the cluster population in M51 is real, and not due to any artifact
or bias in age-dating.

By performing a linear fit to the data on either side of the
enhancement (the age bins for the fit were chosen from the synthesized
cluster population) and comparing the predictions of this linear fit
with the actual data, we estimate that there are more clusters than
one would expect in the case of constant cluster formation by a factor
of 2.2--2.5.  In our M101 dataset, an identical test reveals no
evidence for enhanced cluster formation.

We note that some fraction of the clusters in our sample which have
only \textit{BVI} filter measurements likely have age estimates which put
them in the age range 100--500~Myr, whereas they are actually younger
objects, as we demonstrated in $\S3.2$ and
Figure~\ref{figure:agecomp}.  Such objects can {\it potentially\/}
affect our conclusions concerning a peak in the formation of clusters
several hundred Myr ago.  However, age dating statistics for Field~1
show that the contamination in the intermediate age peak by younger
objects is only at the $\sim$25--30\% level.  Further, \textit{UBVI} filters
were used for age derivations in Field~1, and the $\sim$100--500~Myr
peak in age is clearly seen for this, our best dataset (see \S4.3).
Assuming a similar contamination fraction overall (in Fields~2, 3, 4),
we find that this will not alter our conclusion concerning peaks
in the cluster age distribution.

\begin{figure}
\centering
%\epsscale{1.}
\includegraphics[width=3.5in]{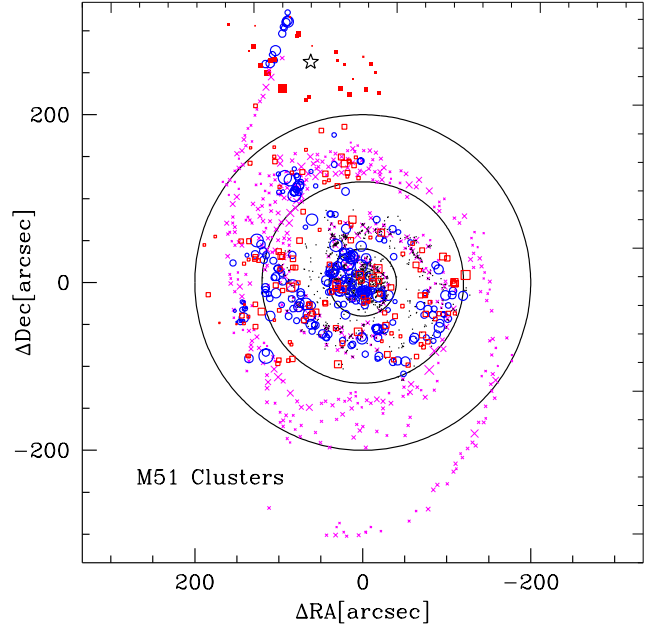}
\caption{The spatial distribution of young clusters (log age $< 8.0$)
in M51 from our sample (open circles), and intermediate age clusters
(log age $\geq8.0$; open squares).  For clusters in field~5 where only
$V$ and $I$ filters are available, we show clusters with $V-I<0.7$ in
blue, and those with $V-I\geq0.7$ in red.  The crosses show H~II regions
on the spiral arm (Rand 1992) and dots mark the locations of H~II regions
in the center of NGC~5194 from Scoville et al.\ (2001).
The open star marks the location of the center of NGC~5195.
Note that the portions which only show H~II regions and no apparent
clusters are due to gaps in the coverage of the WFPC2 pointings (see
Figure~\ref{figure:fov} for the WFPC2 pointings used in this work).
\label{figure:yspatial}}
\end{figure}

Figure~\ref{figure:yspatial} displays the spatial distribution of the
young clusters ($\mbox{log(age[yr])} <8.0$) and intermediate-age
clusters ($\mbox{log(age[yr])} \geq 8.0)$.  Field~5 only has $V$~and
$I$~band imaging, and in this field we consider young clusters to have
colors $(V-I)<0.7$.  In this Figure we also plotted, for comparison,
the H~II regions given by \citet{ran92a} and
\citet{sco01}.  \citet{ran92a} presented a list of H~II regions in the
entirety of M51 based on ground-based observations, while
\citet{sco01} presented a list of H~II regions in the central region
of M51 based on \textit{HST} WFPC2/NICMOS observations.
In the central region (at $r<40$ arcsec) and spiral arms, even given
the incompleteness of our sample, young clusters dominate the sample.
Most young clusters outside the central region are located
along the spiral arms, co-spatial with H~II regions.  Note that the
blue (young) clusters in Field~5 are located right on the extension of
the H~II region spiral arm extending from NGC~5194 passing through
east of NGC~5195.  The absence of ionized gas in this area \citep[see also
the $H\alpha$ maps in][]{thi00, thi02} suggests that these clusters are 
older than typical H~II regions (i.e., older than 5--10~Myr).  

On the other hand, intermediate-age clusters are found both on and off
the spiral arms of NGC~5194. A small number of these clusters are
located in the central region, and they also follow the spiral arms,
but more loosely than the young objects.  In the central region where
a large number of clusters appear to have formed recently, either few
intermediate-age clusters formed, or the majority of them were rapidly
disrupted.  Meanwhile, in the inner-arm region (at $40<r<120$ arcsec),
intermediate age clusters (100--500~Myr) are clearly seen, in
addition to very young clusters.  As found by Bastian et~al.\ (2005),
there appear to be a significantly larger number (by a factor $\sim$2)
of intermediate age clusters on the west side of NGC~5194 relative to
the east side.

\subsection{Cluster Formation Associated with the Interaction of NGC~5194/95?}

The availability of cluster ages allows us to investigate the history
of cluster formation in M51, and compare with dynamical age estimates
of the interaction between NGC~5194 and NGC~5195.  There have been
numerous studies of the dynamical modelling of M51 since
\citet{too72}'s seminal paper explaining galactic bridges and tails in
four interacting galaxies.  To date there are two classes of dynamical
models which explain the morphological structure and dynamics of the
M51 system: nearly parabolic single-passage models 
\citep[e.g.,][]{too72} and bound multiple-passage models \citep{sal00a,
sal00b, dur03}.

In the single-passage models first introduced by \citet{too72}, the
companion galaxy NGC~5195 crossed the NGC~5194 disk in the south at a
distance of 25--30~kpc about 300--500~Myr ago \citep{her90, sal00a,
dur03}.  The model uses an adopted mass ratio of NGC~5195/NGC~5194 of
1/2 to 1/3, consisent with observations \citep{sch77,smi90}.  In this
model NGC~5195 is currently located $\sim$50~kpc behind the M51 disk,
and is moving away from NGC~5194 rapidly ($\Delta v \sim 150$ km~s$^{-1}$).  
The main spiral structures in M51 were formed during the crossing process.

In the multiple-passage model introduced by \citet{sal00a},
the first crossing of the companion is similar to that of the
single-passage model, except the direction of crossing is opposite to
that of the single-passage model (i.e., the companion crosses the
NGC~5194 disk coming toward the observer).  The companion then crossed
the NGC~5194 disk a second time in the north at a distance of 20--25~kpc 
about 50--100~Myr ago.  In this model, NGC~5195 is currently
located less than 20~kpc behind NGC~5194.

The multiple-passage models successfully explain several
observational features of M51, including the detailed HI kinematics
\citep{rot90, sal00a}.  However, \citet{dur03} found, from a
spectroscopic study of the kinematics of the planetary nebula system,
that the kinematics of the north-western tidal tails in M51
can be better explained by the single-passage models, while
the receding component in the north-western tidal tails is consistent
with the results of the multiple-passage model given by \citet{sal00a}
(see their Figure~5).  Therefore it is not yet clear which of these
two models is a better representation of the M51 system.

The age distribution of clusters in NGC~5194
(Figure~\ref{figure:dndt}) clearly shows evidence for a broad peak in
the number of clusters formed at intermediate ages (between 8.0 and
9.0 log yrs), consistent with the predictions of both single- and
multiple- passage models.  However, as noted in the previous section,
rather than a single broad peak in this age range, there may be two
peaks, as observed near ages of 8.0 and 8.6 log(age[yr]).  In order to
assess the robustness of the possible peaks near 100~Myr and 500~Myr,
we performed two experiments.  First, we tested the cluster age
distribution between log(age[yr]) 7.6 and 9.2 for bimodality using the
KMM algorithm (McLachlan \& Basford 1988; Ashman et~al.\ 1994).  As
input, we used the ages for clusters more massive than log mass of 4.5
(as shown in the bottom panel of Figure~\ref{figure:dndt}).  The
p-value returned by KMM for a given distribution measures the
statistical significance of the improvement in the fit when going from
a single gaussian to two gaussians.  The software finds that two peaks
are preferred at the $>99$\% confidence level, with peak values of 8.0
(100~Myr) and 8.7 (500~Myr) log(age[yr]).  In a second test, we ran
Monte Carlo simulations to assess how frequently a peak of the {\it
strength\/} observed in the 8.0 log(age[yr]) bin (shown in
Figure~\ref{figure:dndt}) results from random statistics.  We find
that the peak is real at the $\sim$80\% confidence level.  In general,
while these results are suggestive of two peaks in the age
distribution, one around 100~Myr and the other around 500~Myr, the
results are more sensitive to the accuracy of the age dating and
potential artifacts similar to the 10--30~Myr artifact discussed in
\S3.1.  Given that a number of fields used in this study only have
\textit{BVI} observations which leads to age estimates with relatively large
uncertainties, we conclude that higher quality data is required to
definitively establish whether there is a single or double peak in the
cluster age distribution between 100--500~Myr.  In general however,
the broad peak in the number of clusters formed several hundred Myr
ago is consistent with a scenario where the interaction between the
two galaxies directly caused an increase in the cluster formation rate.

\subsection{Properties of Clusters in NGC~5195}

\begin{figure}
\centering
%\epsscale{1.}
\includegraphics[width=3.5in]{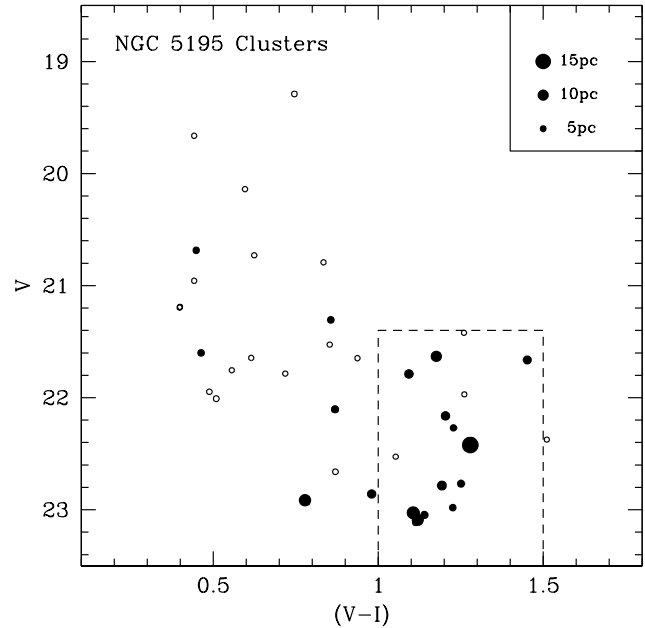}
\caption{$V$ vs. $V-I$ diagram of clusters detected in NGC~5195.
Clusters which have measured sizes less than 5~pc are shown as the
smallest filled circles, clusters with sizes between $5-10$~pc as the
intermediate size circles, and clusters with sizes between $10-15$~pc
as the largest circles.  The dashed lines mark the approximate
color and magnitude boundary found for faint fuzzy star clusters in
NGC~1023 (Larsen \& Brodie 2000).  The majority of
the detected clusters within this box have $r_{eff}$ measurements
of 7~pc and greater.
\label{figure:cmd}}
\end{figure}

Figure~\ref{figure:cmd} shows the color-magnitude diagram (CMD) of all
the clusters we detected in NGC~5195, an SB0 galaxy.  In this diagram,
the size of the symbols represents the relative sizes of the clusters.
The red clusters with $1.0<(V-I)<1.5$ are fainter than $V\approx 21.4$
mag ($M_V\approx -8.3$ mag), while the blue clusters with
$0.4<(V-I)<1.0$ are bright, with $V$ up to $V\approx 19.3$~mag
($M_V\approx -10.5$~mag).  The distribution of the red clusters in
Field~5 is entirely different from that of the blue clusters in the
same field, as seen in Figure~\ref{figure:yspatial}.  The red clusters
are scattered over the face of NGC~5195, while the blue clusters are
located mostly along the arm of NGC~5194 touching NGC~5195.  This
difference indicates that these red clusters are part of a different
population than the blue clusters.  Although with only $V-I$ it is not
possible to establish definitively whether the red clusters in
NGC~5195 are young and highly extincted, or whether they are red
because they are ancient, we suggest that based on the different
distributions of red and blue clusters,
these are ancient star clusters.  Similarly, studies of nearby
lenticulars have established that these galaxies contain almost
exclusively ancient ($\gea \mbox{several}$~Gyr) star clusters (e.g.,
NGC~1023; Larsen \& Brodie 2000).  

In Figure~\ref{figure:cmd}, the sizes of the points reflect their
measured sizes.  It is seen in Figure~\ref{figure:cmd} that the red
clusters with $(V-I)>1.0$ are systematically larger than the blue
clusters.  About 70\% (12 out of 17) red clusters have $r_{eff}$
values larger than 7~pc, while most of the blue, presumably young,
clusters are smaller than 7~pc.  A study of the lenticular (SB0)
galaxy NGC~1023 revealed a new family of star clusters, so-called
``faint fuzzies'' (Larsen \& Brodie 2000).  To date, four early-type
galaxies (three S0s and one E) were searched for faint fuzzy
clusters, and only found in one additional target (NGC~3384), which is
also a barred S0 galaxy \citep{bur05}.  The main characteristics of
these clusters are that they are larger than normal old globular
clusters, with $r_{eff}$ between 7--15~pc; they tend to be faint with
most having $M_V<-7$ mag; and they are red, or metal-rich, with
[Fe/H]~$=-0.58\pm0.24$~dex.  The dashed box in Figure~\ref{figure:cmd}
marks the region of color and luminosity space occupied by faint fuzzy
star clusters.  We conclude that NGC~5195 is the third known SB0
galaxy, after NGC~1023 and NGC~3384, where ancient fuzzy star clusters
have formed and survived.

Although there are relatively few fuzzy clusters detected in NGC~5195
(deeper observations would presumably reveal a larger population), we
can compare their general properties with those found in NGC~1023.
Our sample contains 12 red clusters with $r_{eff}\geq7$~pc, brighter
than $V\sim-6.5$.  Counting the circles in Figure~8 of
\citet{lar00bb}, we estimate that there are a roughly comparable
number of clusters in NGC~1023 to the same absolute magnitude.  The
total cluster magnitudes are dependent on the technique used for
aperture corrections.  We have used a similar technique for making
aperture corrections, so the measured luminosities for clusters in
NGC~1023 should be directly comparable to those presented here for
NGC~5195.  NGC~1023 has a total V band luminosity which is roughly 
1~magnitude brighter than NGC~5195, implying that NGC~1023 is
$\sim$$4\times$ more massive than NGC~5195.  This assumes that both
galaxies are dominated by stellar populations with similar ages;
however the presence of strong Balmer lines in NGC~5195 indicates
recent star formation, which would decrease the mass to light ratio
and lead to an even larger mass difference between the two galaxies.
Therefore, scaled for luminosity, the total number of clusters with
$r_{eff}\geq7$~pc and brighter than $V=-6.5$ appears to be larger in
NGC~5195.  Given the small number of detected red clusters, we were
unable to establish whether the compact ($r_{eff}<7$~pc) and extended
($r_{eff}\geq7$~pc) have different spatial distributions, as found in
NGC~1023 \citep{lar00bb}.  However, deeper observations with \textit{HST}/ACS
may establish whether the faint fuzzy star clusters in NGC~5195 are
also distributed in a ring.

\section{Summary and Conclusions}

We present a study of star clusters in a nearby pair of interacting
galaxies NGC~5194/95 (M51) based on \textit{HST} WFPC2 multi-band archive
images.  We selected a clean cluster sample by only including clusters
with robust morphological information.  This resulted in the detection
of 392 resolved, relatively isolated clusters in five \textit{HST} fields.
However due to this isolation criterion, we estimate that we are
missing the majority (by factors of 4--6) of young clusters
($\lea10$~Myr).  Therefore, in this work we focused on the M51 cluster
population older than 10~Myr.

An age distribution shows a broad enhancement in the number of
clusters with ages between 100~Myr and 500~Myr.  We quantify this
over-density, and estimate that our sample contains a factor of
2.2--2.5 more clusters than expected if the cluster formation rate
had been constant over the past 1~Gyr.  This range of ages is
consistent with the crossing time of the companion galaxy, NGC~5195,
through the NGC~5194 disk.  We see tentative evidence for the presence
of narrower peaks at 100~Myr and 500~Myr in the age distribution.
Although these are consistent with the predictions of multiple-passage
dynamical models of this two-galaxy system, the result is very
preliminary and should be verified.  We estimated the sizes for the
clusters in our M51 sample, and find a correlation between cluster
mass and size at the $\sim4\sigma$ level, and a correlation between
cluster age and size at the $\sim3\sigma$ level.  Finally, we report
for the first time the discovery of faint, extended red star clusters in
NGC~5195, which makes this the third known SB0 galaxy to have formed
``faint fuzzies.''

\acknowledgements

M.G.L.\ acknowledges the support of the collaborative visiting program
of the STScI during this work.  M.G.L.\ thanks the staff of the
Department of Terrestrial Magnetism, Carnegie Institution of
Washington, for their kind hospitality.  M.G.L.\ was supported in part
by the ABRL (R14-2002-058-01000-0) and the BK21 programs.  The authors
are grateful to Minsun Kim for her help during this work, to Richard
Rand and Nick Scoville for providing the lists of M51 HII regions.  We
thank H.~Lamers for making his cluster fitting code available to us.
R.C.\ is grateful for support from NASA through grant GO-09192.01-A
from the Space Telescope Science Institute, which is operated by the
Association of Universities for Research in Astronomy, Inc., for NASA
under contract NAS5-26555.  Finally, we wish to thank the referee,
S.~Larsen, for carefully reading the paper and making a number of
helpful suggestions.
%\clearpage

%\clearpage

%Table 1
\begin{deluxetable}{cccccccc}
\tablecaption{SUMMARY OF HST WFPC2 FIELD OBSERVATIONS IN M51}
\tabletypesize{\footnotesize}
\tablehead{\colhead{Field} &\colhead{PropID} &\colhead{RA} &\colhead{Dec} &\multicolumn{4}{c}{Filters and Exposure Times [s]}\\
\colhead{\#}  &\colhead{}    &\colhead{(J2000)}  &\colhead{(J2000)}  &\colhead{$U$} &\colhead{$B$} &\colhead{$V$} &\colhead{$I$}
}
\startdata
1 &7375 &13:29:48.6 &47:11:30 &F336W, $2\times600$ &F439W, 500,600 &F555W, $2\times600$  &F814W, 300,700\\
2\rlap{\tablenotemark{a}} &5777 &13:29:56.5 &47:11:32 &F336W, $3\times400$\rlap{\tablenotemark{b}} &F439W, $2\times700$ 
&F555W, $1\times600$ &F814W, $1\times600$\\
3 &9073 &13:30:05.4 &47:11:23 &\nodata  &F450W, $4\times500$ &F555W, $4\times500$  &F814W, $4\times500$\\
4 &9073 &13:29:58.7 &47:14:01 &\nodata  &F450W, $4\times500$ &F555W, $4\times500$  &F814W, $4\times500$\\
5\rlap{\tablenotemark{c}} &9042 &13:29:59.0 &47:16:06 &\nodata &\nodata &F606W, $2\times230$ &F814W, $2\times230$\\
\enddata
\tablecomments{Units of right ascension are hours, minutes, and seconds,
and units of declination are degrees, arcminutes, and arcseconds.}
\tablenotetext{a}{Field 2 also has a single F675W (R band) exposure of 600 seconds.}
\tablenotetext{b}{The $U$ band observations for Field 2 were taken at a somewhat different orientation and pointing.
The overlap region is approximately one Wide Field CCD (Chip 2).}
\tablenotetext{c}{Field 5 covers the barred spiral companion galaxy, NGC~5195.}
%\normalsize
\end{deluxetable}

%\clearpage

\end{document}